\documentclass[preprint]{aastex}

\begin{document}

\title{The Solar-Type Contact Binary BX Pegasi Revisited}
\author{Jae Woo Lee, Seung-Lee Kim, Chung-Uk Lee, and Jae-Hyuck Youn}
\affil{Korea Astronomy and Space Science Institute, Daejeon 305-348, Korea}
\email{jwlee@kasi.re.kr, slkim@kasi.re.kr, leecu@kasi.re.kr, jhyoon$@$kasi.re.kr}

\begin{abstract}
We present the results of new CCD photometry for the contact binary BX Peg, made during three successive months 
beginning on September 2008. As do historical light curves, our observations display an O'Connell effect and
the November data by themselves indicate clear evidence for very short-time brightness disturbance. 
For these variations, model spots are applied separately to the two data set of Group I (Sep.--Oct.) and 
Group II (Nov.). The former is described by a single cool spot on the secondary photosphere and the latter by 
a two-spot model with a cool spot on the cool star and a hot one on either star. These are generalized manifestations 
of the magnetic activity of the binary system. Twenty light-curve timings calculated from Wilson-Devinney code 
were used for a period study, together with all other minimum epochs. The complex period changes of BX Peg can be 
sorted into a secular period decrease caused dominantly by angular momentum loss due to magnetic stellar wind braking, 
a light-travel-time (LTT) effect due to the orbit of a low-mass third companion, and 
a previously unknown short-term oscillation. This last period modulation could be produced either by a second LTT orbit 
with a period of about 16 yr due to the existence of a fourth body or by the effect of magnetic activity with 
a cycle length of about 12 yr. 
\end{abstract}


\section{INTRODUCTION}

For BX Peg, Lee at al. (2004a, hereafter Lee04a) remains the most recent comprehensive study. These authors showed 
that historical light curves of BX Peg, displaying year-to-year light variability, can all be explained by introducing 
a single dark spot on the more massive secondary star and found that the orbital period has varied due to 
a periodic oscillation overlaid on a continuous period decrease. They concluded that the periodic $O$--$C$ residuals 
could not be produced by spot activity (Maceroni \& van't Veer 1994) and were not locked into the light variation as 
required by Applegate (1992). Thus, the only phenomenon that could be responsible was a third body with a period of 
52.4 yr and a limiting mass of $M_3 \sin i_3$=0.26 $M_\odot$ and the hypothetical companion became the default explanation.

Lee04a also suggested that the timing residuals indicated an additional short-term oscillation with a period of 
about 12 yr and a semi-amplitude of about 0.002 d. Since then, many new photoelectric and CCD timings of minimum light 
have been reported and should now be sufficiently numerous to test this possibility meaningfully. In this article, 
we present a detailed study of the $O$--$C$ diagram of BX Peg together with a new light-curve synthesis.

\section{NEW PHOTOMETIC OBSERVATIONS}

New CCD photometry of BX Peg was performed on 9 nights from 26 September through 15 November 2008. The observations 
were taken with a 2K CCD camera and a $BVR$ filter set attached to the 1.0-m reflector at Mt. Lemmon Optical Astronomy 
Observatory (LOAO) in Arizona, USA. The instrument and reduction method have been described by Lee et al. (2009). 
The comparison star (C) was chosen to be ${\rm BD}+25^{\rm o}4584$ used in the previous observations by 
Zhai \& Zhang (1979), Samec (1990), and Lee04a, all of whom reported no light variations for it. 2MASS 21392958+2637157
was selected as a check star (K) to verify the constancy in brightness of the comparison star. The reference stars 
were imaged on the CCD chip at the same time as the program star. 

A total of 1,235 individual observations was obtained among the three bandpasses (411 in $B$, 412 in $V$, and 412 in $R$) 
and a sample of them is listed in Table 1. The light curves of BX Peg are plotted in the upper panel of Figure 1 and 
the (K--C) magnitude differences are shown in the lower panel. Measurements of the check star indicate that, on average, 
the comparison star remained constant within the 1$\sigma$-value of about $\pm$0.006 mag during our observing runs.

\section{LIGHT-CURVE SYNTHESIS}

Our light curves of BX Peg are asymmetrical and continue to display season-to-season light variability as they have in
previous years. From the analysis of historical data, Lee04a showed that the asymmetries can be interpreted as spot activity 
on the secondary component presumably produced by a magnetic dynamo, and the variations of the asymmetries most likely 
arise from the variability of the spots with time. As shown in Figure 1, the new light curves still indicate an 
O'Connell effect (Max I fainter than Max II) and cycle-to-cycle intrinsic variability. Specifically, the November light 
curves are very different from the other data, especially at the first quadrature. We solved the new light curves 
in a manner almost identical to that used by Lee04a. For this synthesis, the contact mode 3 of 
the 2003 version\footnote {ftp://ftp.astro.ufl.edu/pub/wilson/} of the Wilson-Devinney code 
(Wilson \& Devinney 1971; Wilson 1979, 1990; hereafter WD) was applied separately to the two data sets of 
Group I (Sep.--Oct.) and Group II (Nov.) and a cool spot on the secondary photosphere was adopted for both Groups.  

To start, we analyzed the light curves of Group I as reference ones. The photometric solutions are given in Table 2 
and the spot parameters in the second column of Table 3, wherein the primary refers to the less massive hot star 
and the secondary to the more massive cool one. The results appear as the continuous curves in Figure 1 and the Group I 
and Group II residuals from this spot model are plotted in the left and middle panels of Figure 2, respectively. 
The single cool spot on the secondary represents the light curves of Group I satisfactorily, but a second spot is needed 
to explain the light residuals for Group II. Therefore, we modelled Group II data by adding a hot spot on the cool star 
and by adjusting only the spot and luminosity characteristics among the model parameters. These results are listed 
in columns (3)-(4) of Table 3 and the residuals from the model are plotted in the right panel of Figure 2, 
where we see that the hot spot is sufficient to explain the residual light discrepancy. 

A separate trial, given in columns (5)-(6) of Table 3, testing for a hot spot on the primary star, was as successful 
as for the hot spot on the secondary so it is not possible to discriminate between these two possibilities. 
This troublesome degeneracy is due to the high inclination of the orbit and the radius difference between the stars 
causing such a hot spot to be visible throughout the same phase interval no matter to which star it is assigned. Lastly, 
we looked for a possible third light source ($\ell_{3}$) but found that the code always returned negative values 
for this parameter.

In Lee04a it was shown that, within errors, 10 independent light curves obtained from 1978 through 2000 could all be 
represented by a unique geometry and by wavelength-consistent photometric parameters.  That conclusion can now be 
extended to encompass the three 2008 light curves reported here.  One must understand that the formal errors returned 
by the WD code are lower limits to realistic uncertainties but, even so, the agreement over 31 years is strong enough 
that the component stars of BX Peg are well known even within the formal errors.  Perhaps this is not too surprising 
since the eclipses are complete and thus light curve determinacy is high, but the implication of the statement is 
that further photometric interest in this binary may be limited to mapping its intrinsic variability.

With this advantage, it is possible to ask whether there has been stability to the cool spots that have been described 
over the same time interval?  Within assigned errors, these spots have migrated to larger longitudes over almost 
a hemisphere and (non-monotonically) have moved closer to the stellar equator.  The spot sizes reached a maximum 
around 2000 and they have cooled progressively over the interval of monitoring. Of course, these conclusions depend 
on light curves that under-sample the time interval over which they were accumulated.  It is also clear that a hot spot 
can emerge in a very brief time as happened over only a month in 2008.  Almost certainly, this means that the hot spot 
is not a signature of impact from streaming gas since the over-contact condition has been documented since 1978. Rather, 
the hot disturbance and the variability of the cool spots are generalized manifestations of the magnetic activity 
of the BX Peg system.

\section{FITTING THE $O$--$C$ VARIATION}

We calculated minimum epochs for each of our eclipses with the WD code by means of adjusting only the ephemeris epoch 
($T_0$). Ten such timings of minimum light are given in Table 4, together with all photoelectric and CCD timings 
since the compilation of Lee04a. For further ephemeris improvement, we used the light-curve timings given in Table 3 
of Lee04a, rather than the original minima of Samec (1990). The following standard deviations were assigned to 
timing residuals based on observational technique and the method of measuring the epochs: $\pm$0.0036 d for visual 
and photographic, $\pm$0.0009 d for photoelectric and CCD, and $\pm$0.0003 d for WD minima. Relative weights were then 
scaled from the inverse squares of these values. Lee04a has shown that orbital period changes of BX Peg can be represented 
by a quadratic {\it plus} LTT ephemeris:
\begin{eqnarray}
C_1 = T_0 + PE + AE^2 + \tau_3,
\end{eqnarray}
where $\tau_3$ symbolizes the LTT due to a third companion physically bound to the eclipsing pair (Irwin 1952, 1959) 
and includes five parameters ($a_{12}\sin i_3$, $e$, $\omega$, $n$ and $T$). In order to improve the coefficients of
the former ephemeris,  we fitted all times of minimum light to equation (1) using 
the Levenberg-Marquart (LM) algorithm (Press et al. 1992). The results are given in the second column of Table 5, 
together with related parameters. The absolute dimensions of Samec \& Hube (1991) have been used for these and 
subsequent calculations. The value for the third-companion period is different from the evaluation in Lee04a because  
the intervening years have added much weight to the period history.  The other Irwin parameters have not changed 
significantly if one recognizes that $\omega$ is poorly determined even now. 

The top panel of Figure 3 shows the $O$--$C_1$ residuals constructed with the linear term of the ephemeris; 
the solid curve and the dashed parabola represent the full non-linear terms and the quadratic term, respectively. 
The middle panel displays the residuals $\tau_3$ from the linear and quadratic terms of the equation and 
the bottom panel the residuals from the complete ephemeris. These appear as $O$--$C_1$ in the third column of Table 4. 
In all panels, error bars are shown for only the timings with known errors. In the bottom panel of Figure 3, 
the timing residuals indicate, as before, a possible additional short-term oscillation. 
Accordingly, the period variability of the system must be more complicated than the form of equation (1). 
To get a more generalized description of the period variability, we introduced the times of minimum light into 
a different ephemeris form:
\begin{eqnarray}
C_2 = T_0 + PE + AE^2 + \tau_3 + \it {S}, ~~\it {S} = K^\prime \sin(\omega^\prime E + \omega_0 ^\prime).
\end{eqnarray}
The LM technique was again applied to solve for the eleven parameters of the ephemeris which are listed in the third column 
of Table 5. Adding the generalized sine modulation decreased the third-body period to a value close to that in Lee04a
but nothing else has changed. The $O$--$C_{2}$ residuals from the linear light elements are plotted in the top panel of 
Figure 4. The second and third panels display the LTT orbit ($\tau_3$) and a 12-yr period modulation ($\it S$), respectively, 
and the lowest panel the residuals from the full equation (2). These appear as $O$--$C_2$ in the fourth column of Table 4. 
It is clear that the residuals in the third panel of Figure 4 skew across the sine curve and the fit is not so good as could 
be wished. Consequently, the lowest panel of the figure does not show the final residuals as randomized as they should be.
 
The sine oscillation can be replaced in favor of a second LTT orbit ascribed to a fourth component of the BX Peg system. 
For such a case, it is necessary to use a quadratic {\it plus} two-LTT ephemeris instead of equation (2):
\begin{eqnarray}
C_3 = T_0 + PE + AE^2 + \tau_3 + \tau_4.
\end{eqnarray}
These calculations converged quickly to yield the entries listed in Table 6 where we see that not much has changed 
for the orbit of the supposed third object of the system. Figure 5, derived from the Table 6 parameters, is plotted 
in the same sense as Figure 4. The second and third panels refer to the $\tau_3$ and $\tau_4$ orbits, respectively, 
and are possibly marginally better fits to the data. If the hypothetical objects are on the main sequence, 
the minimum masses for the putative third and fourth bodies correspond to spectral types of M6 and M8, respectively, 
and their bolometric luminosities would contribute only 0.6 \% to the total luminosity of the quadruple system. Also, 
the semi-amplitude of the systemic radial velocity variation of the eclipsing pair due to the two supposed companions 
would be only  about 1 km s$^{-1}$. These two limits indicate that it will be not easy to detect companions 
orbiting the eclipsing binary independently.

As predicted by Maceroni \& van't Veer (1994) and confirmed by Lee et al. (2009), times of minimum light may be 
systematically shifted by light asymmetries due to starspot activity. The light curve synthesis method gives 
more precise timings than do other techniques (e.g. Kwee \& van Woerden 1956) based on the observations 
during a minimum alone.  Clear evidence for this assertion appears in Figures 3--5 which show almost no noise 
for the WD timings compared to those from other methods.

\section{DISCUSSION}

The negative coefficients of the quadratic terms in equations (2) and (3) yield a continuous period decrease with 
a rate of $-$9.8 $\times$10$^{-8}$ d yr$^{-1}$, which can be explained either by conservative mass transfer 
from the more massive cool star to its less massive hot component or by angular momentum loss (AML) due to 
a magnetic stellar wind. Under the assumption of conservative mass transfer, the transfer rate is 
about 7.0$\times$10$^{-8}$ M$_\odot$ yr$^{-1}$. This value is 3.5 times greater than a rate of 
2.0$\times$10$^{-8}$ M$_\odot$ yr$^{-1}$ calculated by assuming that the cool secondary transfers mass to 
the hot primary component on a thermal time scale. Therefore, the alternative mechanism, AML caused by 
magnetic braking rooted in the convective zone, seems more likely. From an approximate formula given 
by Guinan \& Bradstreet (1988), the period decrease rate is calculated to $-$4.1, $-$5.8, and $-$8.7 
(in units of 10$^{-8}$ d yr$^{-1}$) for $k^2$=0.07, 0.10, and 0.15, respectively. The last value might be 
a good approximation to the gyration constant $k^2$ of BX Peg. It is also possible that the correct explanation 
of the secular period change is some combination of non-conservative mass transfer and AML but, at present, 
it seems that AML should be the dominant contributor. 

This last cautionary remark actually has some independent support. Consider the pair of W-type systems defined 
observationally by Binnendijk (1970), V829 Her (Erdem \& \"Ozkarde\c s 2006) and V781 Tau (Yakut et al. 2005). 
Their individual masses and radii and the photospheric temperature difference as well as the systemic mass ratio 
and period are the same values within 1 \% yet the algebraic signs of the secular period changes are different. 
This is not a unique case. The same situation exists for V417 Aql (Qian 2003; Lee et al. 2004b) and BB Peg 
(Kalomeni et al. 2007). If there is little mechanical and radiometric difference between two such binaries 
which behave differently dynamically, there must be a particular evolutionary mechanism that distinguishes them 
or else the seeming secular period changes are not really unbounded and are just long-term cyclical effects. 
The rigorous association of a positive period change with a particular sense of mass transfer and 
the mirror association of a negative period change with the other sense of mass transfer and with AML appear 
to be too didactic.

The 12-yr period modulation, shown in the third panel of Figure 4, could operate in at least one component 
since each has a convective envelope and thus may be magnetically active (Applegate 1992;  Lanza et al. 1998). 
With the values of $K^{\prime}$ and $P^{\prime}$, the parameters of an Applegate model were calculated for both components 
and appear in Table 7, where $\Delta m_{\rm rms}$ denotes a bolometric magnitude difference relative to the mean light level 
of BX Peg converted to magnitude scale with equation (4) in the paper of Kim et al. (1997). The variations of 
the gravitational quadrupole moment $\Delta Q$ correspond to typical values for contact binaries and 
the required light variation associated with each component is within the value ($\Delta L/L_{\rm {p,s}} \sim 0.1$) 
proposed by Applegate. This consistency indicates that Applegate's mechanism could possibly function in both component stars. 
There is no {\it a priori} reason to require a sine-type behavior for a magntic cycle in a star. If Applegate's mechanism 
is the main cause of the cyclical variation, his model requires the brightness variation to vary in phase with 
the period modulation of Figure 4.    

In order to check this possibility and to study the long-term light variations of BX Peg, we measured the light levels 
at four different phases (Max I, Min I, Max II and Min II) for our new light curves and for the archival measurements 
of Zhai \& Zhang (1979), Samec (1990) and Lee04a.  The latter are taken from Table 6 of Lee04a and the comparisons 
are given in Table 8.  All data sets are referred to the same comparison star (${\rm BD}+25^{\rm o}4584$). 
The brightness variations of $\Delta B$ and $\Delta V$ are plotted in Figure 6, where the seasonal means were subtracted 
from the grand mean for all seasons giving the mean seasonal differences in the natural systems. The year and epoch 
for each data set were calculated by averaging the starting and ending HJDs of the observations. In Figure 6, the third 
and fourth panels represent the sine term of the $C_2$ ephemeris and the $\tau_4$ orbit of the $C_3$ ephemeris, 
respectively, averaged over each observing season and the arrows indicate the mean epoch of each seasonal light curve. 
There are only 5 epochs for light curves but, even as few as they are, they do not conform to this prediction 
of the Applegate mechanism. 

This binary presents a conflict for observers.  Without the minimum monitoring which has been so fruitful, we would be 
ignorant of the time-scales and semi-amplitudes of the 50-year cycle and of the shorter one as well.  Perusal of 
the present Figures 3 through 5 and of Figures 4 and 5 in Lee04a shows that the 50-year cycle has become much better 
delineated over only 5 years.  The same conclusion cannot be drawn for the shorter cycle.  We have already remarked that 
neither the sinusoid nor a fourth-companion waveform track the residuals well.  Many of the minimum timings are much less 
precise than those from LOAO and their noisy appearance is readily evident in all the figures.  Residuals as large as 
$-$3.7 minutes appear in Table 4, a value that is almost 1 \% of the Keplerian period and more than 4 \% of 
the entire eclipse width.  Such discrepancies probably do not signify high-frequency spot activity; at least there is 
no independent evidence for such a possibility.  Rather, they must be the result of clerical or observational errors or 
of greatly-undersampled timing determinations.

\acknowledgments{ }
The authors wish to thank Dr. Robert H. Koch for careful readings and corrections and for some helpful comments on 
the draft version of the manuscript. We also thank the staff of Mt. Lemmon Optical Astronomy Observatory for assistance 
with our observations. This research has made use of the Simbad database maintained at CDS, Strasbourg, France.

\clearpage
\begin{figure}
 \includegraphics[]{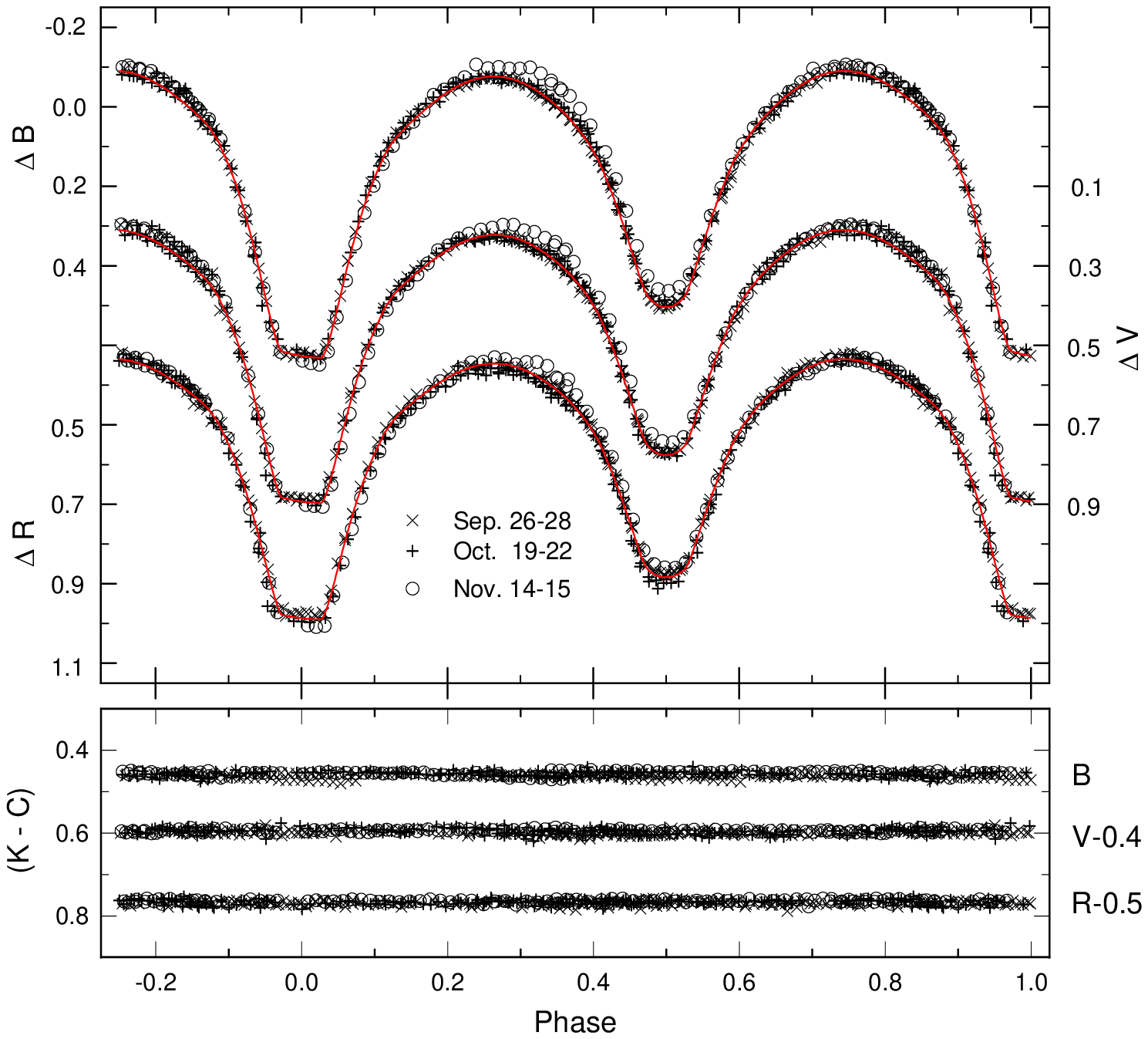}
 \caption{The upper panel displays $BVR$ light curves of BX Peg defined by individual observations. 
 The solid curves represent the photometric solutions obtained from the measurements between September and October 2008. 
 The magnitude differences (K--C) between the check and comparison stars are plotted in the lower panel. }
 \label{Fig1}
\end{figure}

\begin{figure}
 \includegraphics[]{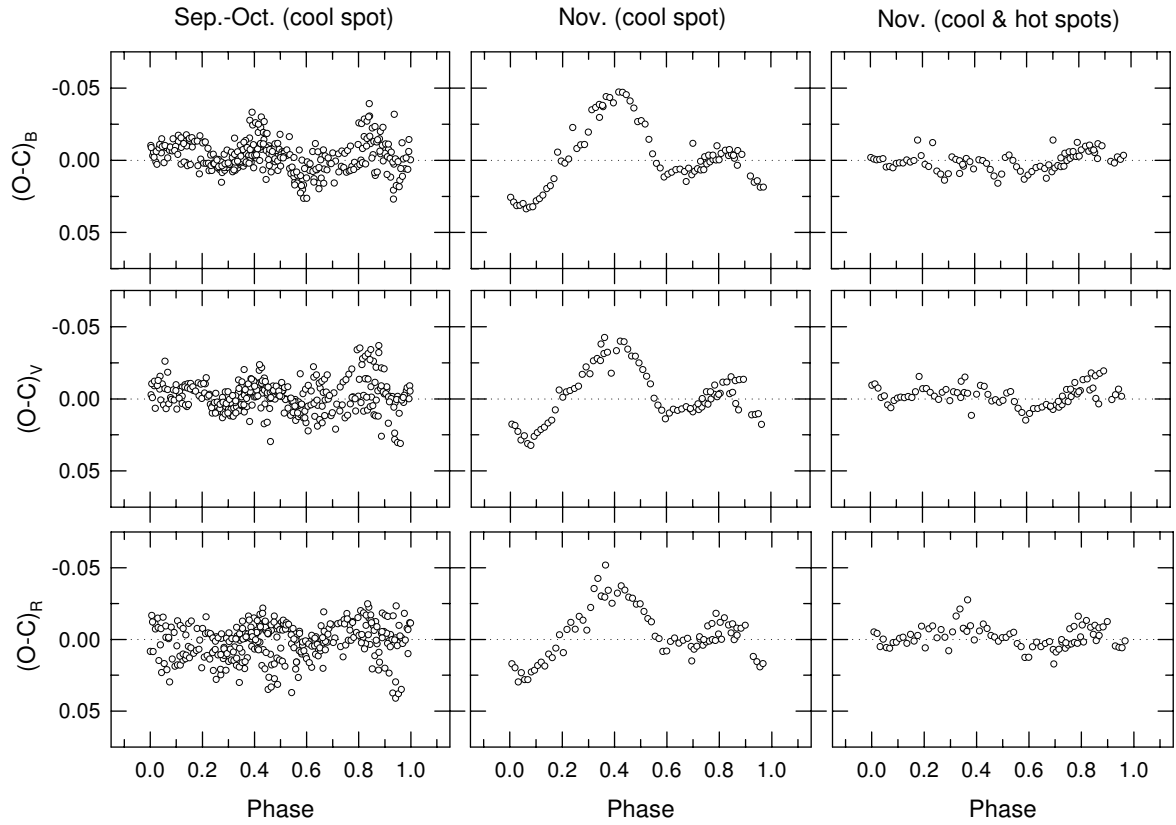}
 \caption{Light residuals from the models for two data sets (Group I and Group II). See the text for details. }
 \label{Fig2}
\end{figure}

\begin{figure}
 \includegraphics[]{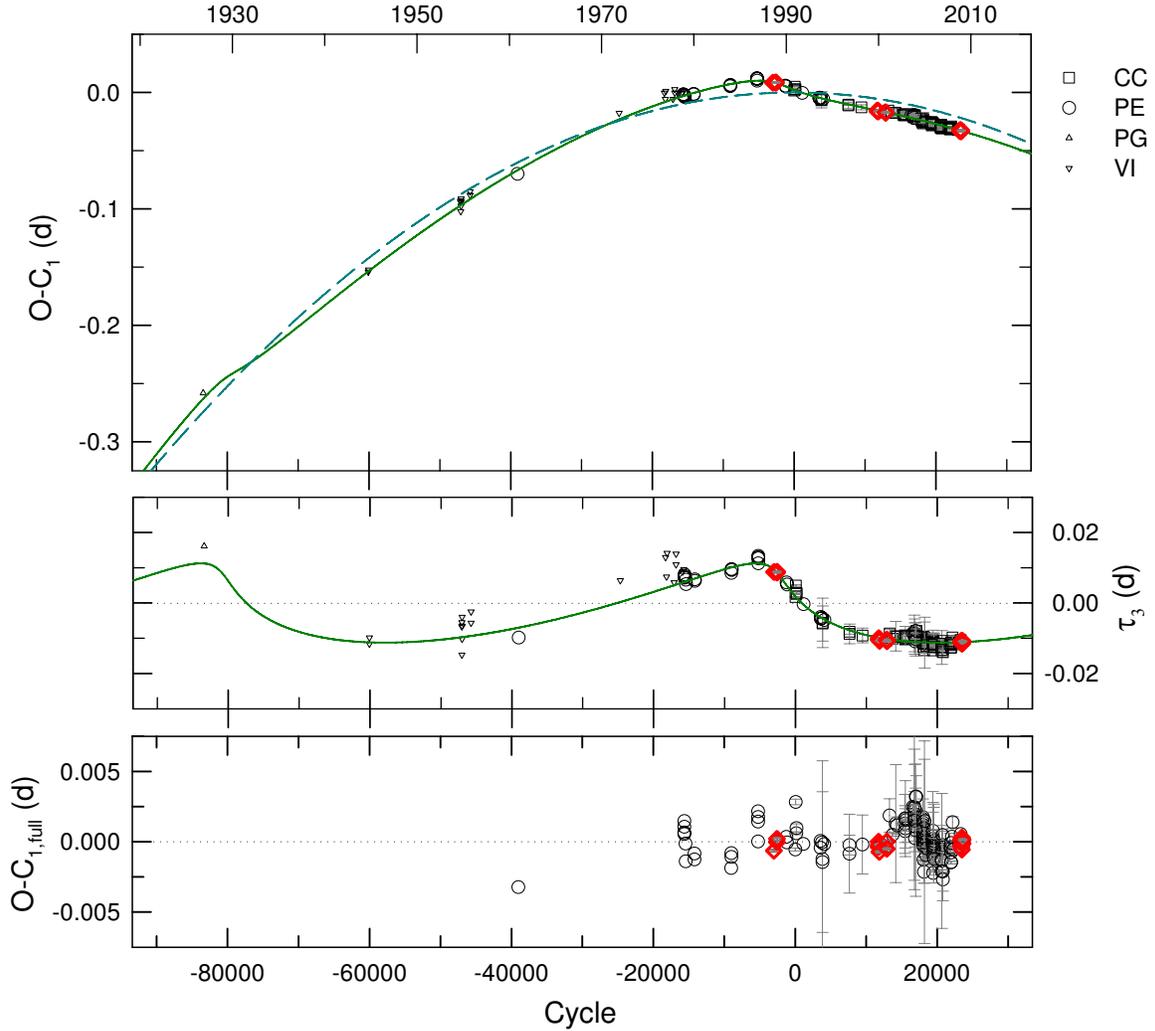}
 \caption{The $O$--$C_{1}$ diagram of BX Peg constructed with the linear terms of the quadratic {\it plus} LTT ephemeris. 
 In the top panel, the continuous curve and the dashed, parabolic one represent the full contribution and the quadratic term 
 of the equation, respectively. Diamond symbols refer to the minimum times obtained with the WD code. The middle panel 
 displays the residuals from the linear and quadratic terms and the bottom panel the residuals from the full ephemeris. 
 An additional short-term oscillation seems to exist in the final residuals in the bottom panel. }
 \label{Fig3}
\end{figure}

\begin{figure}
 \includegraphics[]{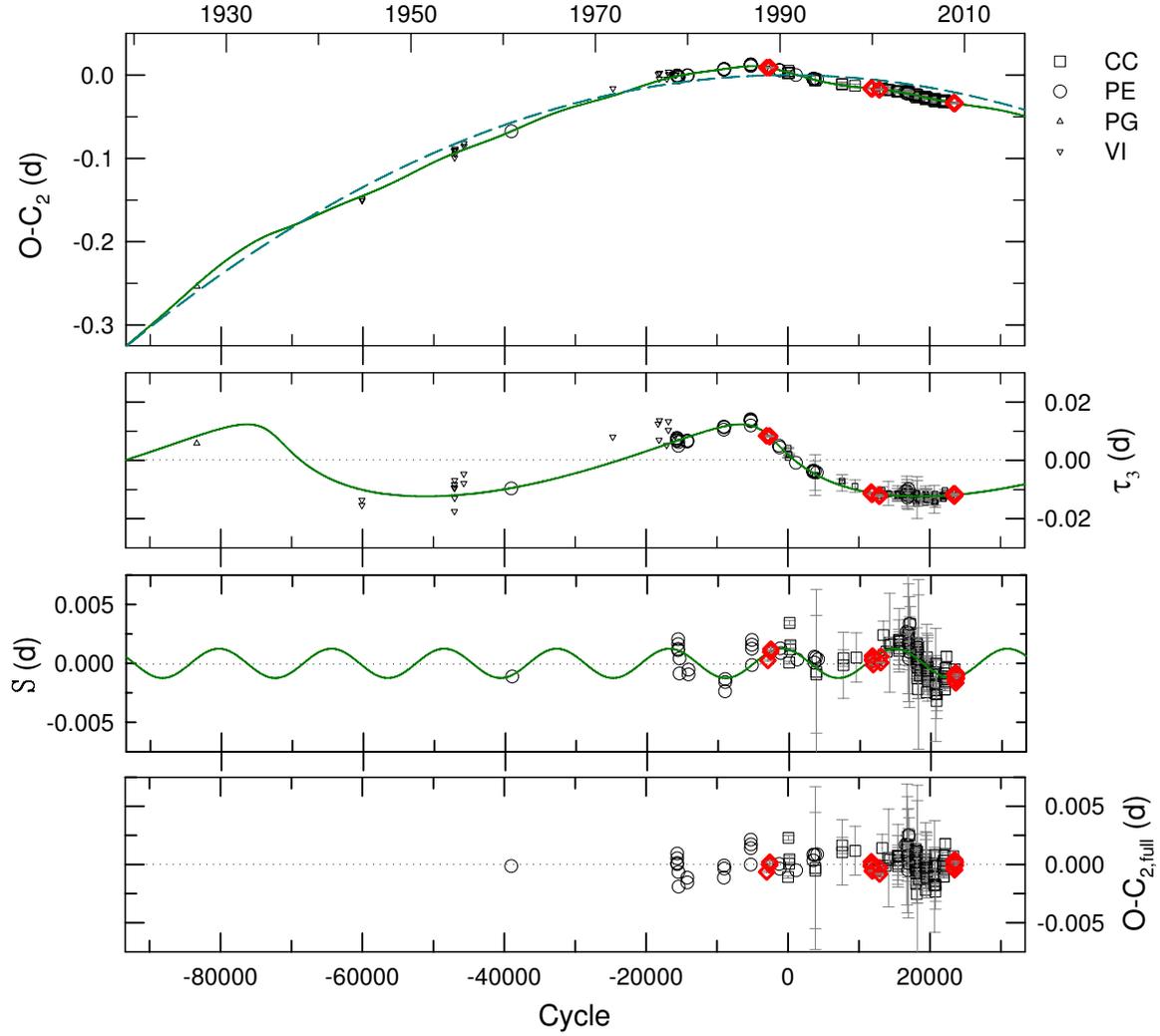}
 \caption{The $O$--$C_{2}$ residuals of BX Peg from the linear ephemeris of equation (2). These are drawn in the top panel
 with the continuous curves due to the full non-linear terms and the dashed parabola due to the quadratic term of the equation.
 The second and third panels display a LTT orbit and a 12-yr period oscillation, respectively, and the lowest panel
 the residuals from the complete equation. }
\label{Fig4}
\end{figure}

\begin{figure}
 \includegraphics[]{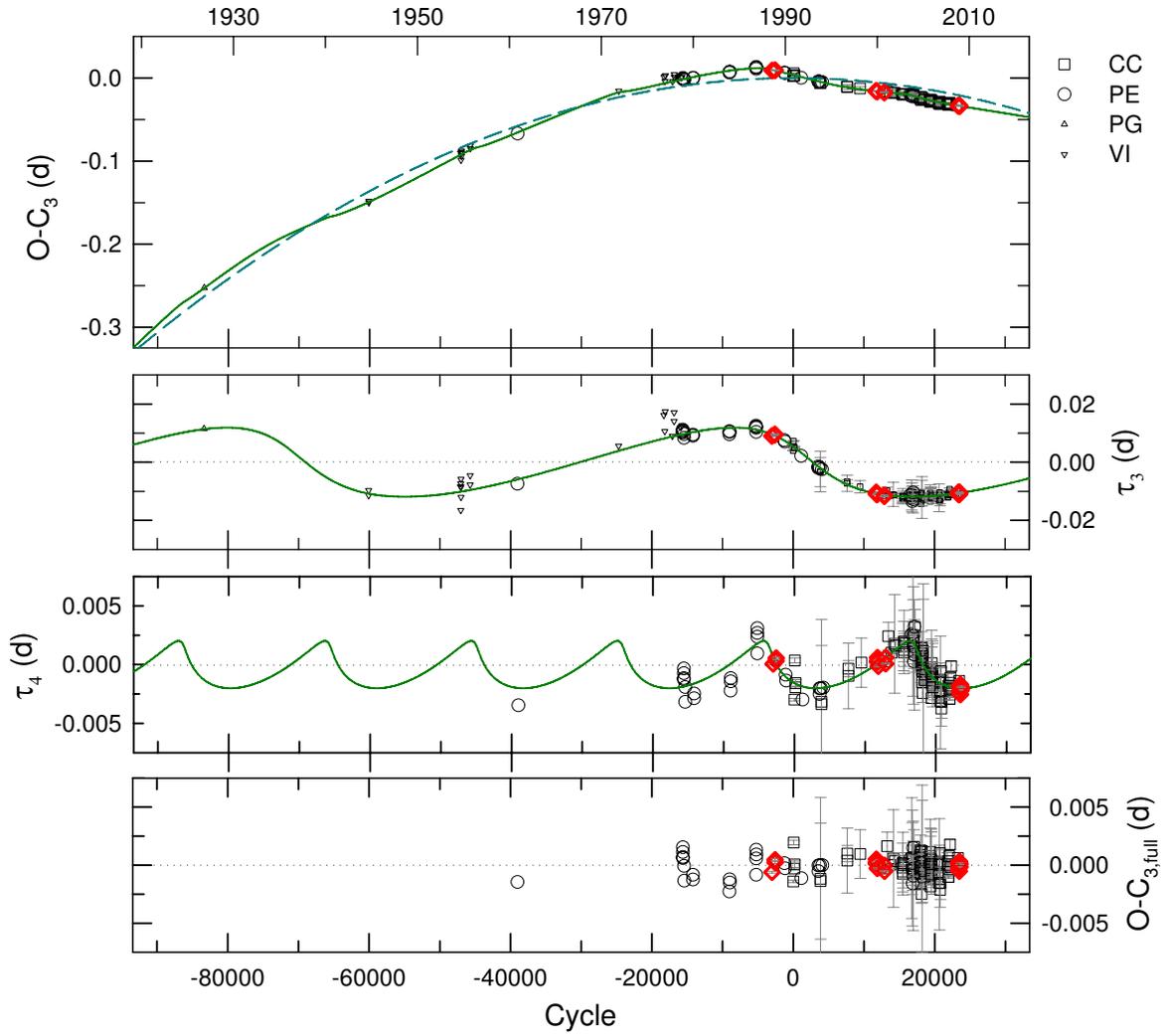}
 \caption{The uper panel is constructed in the same sense as Figure 4 with the linear terms of Table 6. The second and 
 third panels refer to the $\tau_3$ and $\tau_4$ orbits, respectively. }
\label{Fig5}
\end{figure}

\begin{figure}
 \includegraphics[]{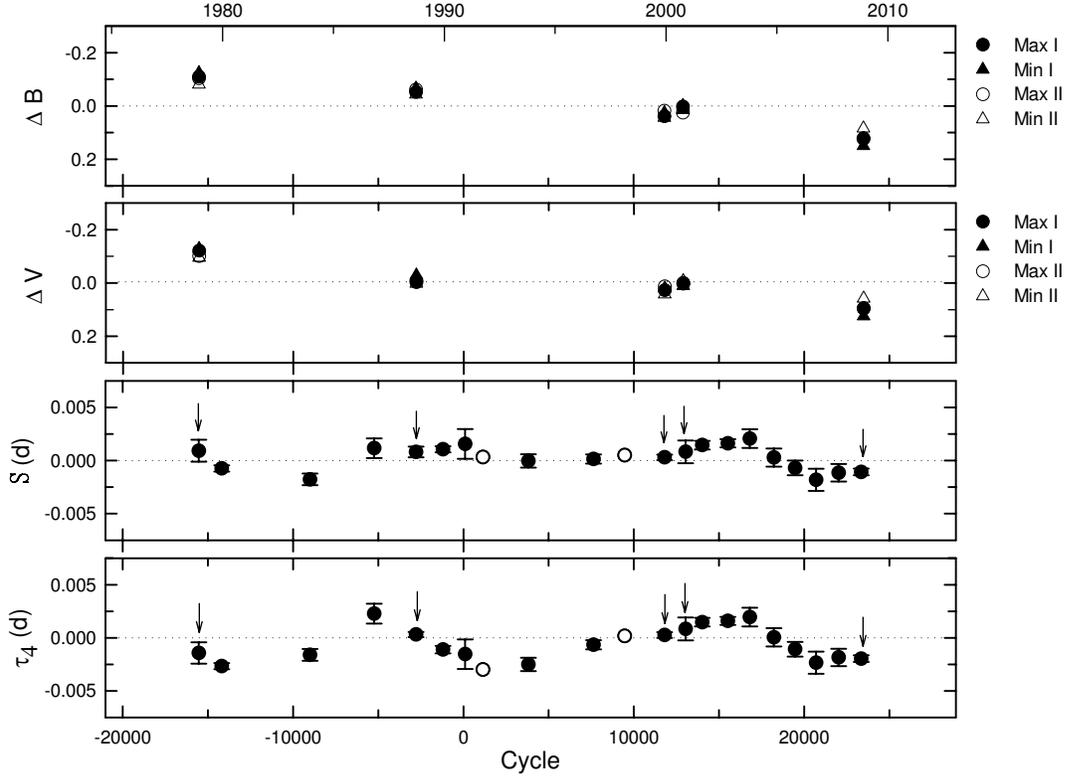}
 \caption{The mean seasonal variations of $\Delta B$ and $\Delta V$ for BX Peg at four characteristic phases 
 (Max I, Min I, Max II and Min II). The third and fourth panels represent the sine term and the $\tau_3$ orbit of 
 the $C_2$ and $C_3$ ephemeris forms, respectively, averaged over each observing season and the plotted points were 
 determined by combining all eclipse timings for a given year. Open circles refer to seasons with only one minimum 
 timing. An errors bar gives the standard deviation of each data set. 
 The arrows indicate the mean epoch of each seasonal light curve. }
 \label{Fig6}
\end{figure}

\clearpage
\begin{deluxetable}{crcrcr}
\tablewidth{0pt} \tablecaption{CCD photometric observations of BX Peg.}
\tablehead{
\colhead{HJD} & \colhead{$\Delta B$} & \colhead{HJD} & \colhead{$\Delta V$} & \colhead{HJD} & \colhead{$\Delta R$} 
}
\startdata
2,454,736.66243 & $-$0.070  &  2,454,736.66347  &  0.227  &  2,454,736.66423  &  0.346   \\
2,454,736.66521 & $-$0.066  &  2,454,736.66625  &  0.231  &  2,454,736.66701  &  0.355   \\
2,454,736.66800 & $-$0.064  &  2,454,736.66904  &  0.235  &  2,454,736.66978  &  0.353   \\
2,454,736.67080 & $-$0.060  &  2,454,736.67190  &  0.242  &  2,454,736.67265  &  0.362   \\
2,454,736.67367 & $-$0.054  &  2,454,736.67477  &  0.250  &  2,454,736.67549  &  0.368   \\
2,454,736.67652 & $-$0.039  &  2,454,736.67761  &  0.259  &  2,454,736.67835  &  0.384   \\
2,454,736.67938 & $-$0.028  &  2,454,736.68047  &  0.270  &  2,454,736.68120  &  0.388   \\
2,454,736.68221 & $-$0.016  &  2,454,736.68331  &  0.286  &  2,454,736.68404  &  0.404   \\
2,454,736.68505 & $-$0.003  &  2,454,736.68614  &  0.297  &  2,454,736.68687  &  0.417   \\
2,454,736.68788 & $ $0.015  &  2,454,736.68898  &  0.318  &  2,454,736.68972  &  0.430   \\
\enddata
\tablecomments{This table is available in its entirety in machine-readable and Virtual Observatory (VO) forms 
in the online journal. A portion is shown here for guidance regarding its form and content.}
\end{deluxetable}

\begin{deluxetable}{ccc}
\tablewidth{0pt}
\tablecaption{Photometric solutions of BX Peg.}
\tablehead{
\colhead{Parameter}       & \colhead{Primary}        & \colhead{Secondary}
}
\startdata
$T_0$ (HJD)               & \multicolumn{2}{c}{2,453,208.3118$\pm$0.0035}               \\
$P$ (d)                   & \multicolumn{2}{c}{0.28041759$\pm$0.00000064}               \\
$q$                       & \multicolumn{2}{c}{2.6897$\pm$0.0045}                       \\
$i$ (deg)                 & \multicolumn{2}{c}{87.693$\pm$0.095}                        \\
$T$ (K)                   & 5532$\pm$20                 & 5300                          \\
$\Omega$                  & 6.135$\pm$0.010             & 6.135                         \\
$A$                       & 0.5                         & 0.5                           \\
$g$                       & 0.32                        & 0.32                          \\
$X$                       & 0.517                       & 0.525                         \\
$x_B$                     & 0.636$\pm$0.049             & 0.662$\pm$0.018               \\
$x_V$                     & 0.440$\pm$0.044             & 0.591$\pm$0.017               \\
$x_R$                     & 0.334$\pm$0.040             & 0.516$\pm$0.018               \\
$L/(L_1+L_2)_B$           & 0.352$\pm$0.001             & 0.648                         \\
$L/(L_1+L_2)_V$           & 0.348$\pm$0.001             & 0.652                         \\
$L/(L_1+L_2)_R$           & 0.341$\pm$0.001             & 0.659                         \\
$r$ (pole)                & 0.2820$\pm$0.0009           & 0.4435$\pm$0.0007             \\
$r$ (side)                & 0.2946$\pm$0.0010           & 0.4752$\pm$0.0009             \\
$r$ (back)                & 0.3310$\pm$0.0016           & 0.5034$\pm$0.0012             \\
$r$ (volume)$\rm ^a$      & 0.3042                      & 0.4755                        \\[1.0mm]
\enddata
\tablenotetext{a}{Mean volume radius.}
\end{deluxetable}

\begin{deluxetable}{cccccccc}
\tablewidth{0pt}
\tablecaption{Spot parameters for BX Peg.$\rm ^a$}
\tablehead{
\colhead{Parameter}       & \colhead{Group I}   && \multicolumn{5}{c}{Group II}                                                    \\ [1.0mm] \cline{4-8} \\ [-2.0ex]
                          & \colhead{Cool 2}    && \multicolumn{2}{c}{Cool 2 and Hot 2} && \multicolumn{2}{c}{Cool 2 and Hot 1}                   
}                                                                                                             
\startdata                                                                                                    
Colatitude (deg)          & 66.6$\pm$2.1        &&  76.2$\pm$1.5     &  61.5$\pm$1.4    &&  78.7$\pm$0.6     &  68.3$\pm$3.2       \\
Longitude (deg)           & 142.7$\pm$0.4       &&  143.7$\pm$2.0    &  40.7$\pm$3.1    &&  140.2$\pm$0.3    &  230.0$\pm$1.1      \\
Radius (deg)              & 14.1$\pm$0.1        &&  15.1$\pm$1.2     &  14.6$\pm$2.0    &&  15.7$\pm$0.1     &  20.7$\pm$0.8       \\
$T$$\rm _{spot}$/$T$$\rm _{local}$  &  0.828$\pm$0.050  &&  0.751$\pm$0.031  &  1.122$\pm$0.005  &&  0.762$\pm$0.025  &  1.117$\pm$0.011  \\
\enddata
\tablenotetext{a}{Cool 2: a cool spot on the secondary; Hot 2: a hot spot on the secondary; Hot 1: a hot spot on the primary.}
\end{deluxetable}

\begin{deluxetable}{lrrrrcl}
\tabletypesize{\scriptsize}
\tablewidth{0pt}
\tablecaption{Observed photoelectric and CCD times of minimum light for BX Peg since the compilation of Lee04a.}
\tablehead{
\colhead{HJD}   & \colhead{Epoch}  & \colhead{$O$--$C_{1}$} & \colhead{$O$--$C_{2}$} & \colhead{$O$--$C_{3}$} & \colhead{Min} & References \\
\colhead{(2,450,000+)} & & & & & & }
\startdata
1,899.3225   &   13283.0   &   0.00186   &    0.00140   &    0.00164   &   I    &   Br\'at et al. (2007)          \\
2,145.5289   &   14161.0   &   0.00129   &    0.00056   &    0.00057   &   I    &   Zejda (2004)                  \\
2,521.4288   &   15501.5   &   0.00095   &    0.00006   &   -0.00056   &   II   &   Zejda (2004)                  \\
2,521.5697   &   15502.0   &   0.00164   &    0.00075   &    0.00013   &   I    &   Zejda (2004)                  \\
2,878.4018   &   16774.5   &   0.00197   &    0.00122   &    0.00012   &   II   &   H\"ubscher et al. (2005)      \\
2,878.5425   &   16775.0   &   0.00246   &    0.00171   &    0.00061   &   I    &   H\"ubscher et al. (2005)      \\
2,886.3935   &   16803.0   &   0.00176   &    0.00102   &   -0.00008   &   I    &   Krajci (2005)                 \\
2,887.3756   &   16806.5   &   0.00240   &    0.00165   &    0.00056   &   II   &   H\"ubscher (2005)             \\
2,887.5149   &   16807.0   &   0.00149   &    0.00074   &   -0.00035   &   I    &   H\"ubscher (2005)             \\
2,887.515    &   16807.0   &   0.00159   &    0.00084   &   -0.00025   &   I    &   Diethelm (2004)               \\
2,902.3758   &   16860.0   &   0.00024   &   -0.00049   &   -0.00157   &   I    &   H\"ubscher (2005)             \\
2,929.4367   &   16956.5   &   0.00082   &    0.00011   &   -0.00092   &   II   &   H\"ubscher (2005)             \\
2,929.5793   &   16957.0   &   0.00321   &    0.00250   &    0.00147   &   I    &   H\"ubscher (2005)             \\
2,956.3592   &   17052.5   &   0.00320   &    0.00251   &    0.00156   &   II   &   Br\'at et al. (2007)          \\
3,208.4530   &   17951.5   &   0.00137   &    0.00092   &    0.00086   &   II   &   Pribulla et al. (2005)        \\
3,209.4332   &   17955.0   &   0.00010   &   -0.00034   &   -0.00040   &   I    &   H\"ubscher et al. (2005)      \\
3,209.4335   &   17955.0   &   0.00040   &   -0.00004   &   -0.00010   &   I    &   Pribulla et al. (2005)        \\
3,209.5743   &   17955.5   &   0.00099   &    0.00055   &    0.00049   &   II   &   Pribulla et al. (2005)        \\
3,212.5180   &   17966.0   &   0.00031   &   -0.00013   &   -0.00018   &   I    &   Pribulla et al. (2005)        \\
3,217.4262   &   17983.5   &   0.00120   &    0.00076   &    0.00072   &   II   &   H\"ubscher et al. (2005)      \\
3,220.3701   &   17994.0   &   0.00071   &    0.00028   &    0.00025   &   I    &   Pribulla et al. (2005)        \\
3,220.5099   &   17994.5   &   0.00030   &   -0.00013   &   -0.00016   &   II   &   Pribulla et al. (2005)        \\
3,220.5112   &   17994.5   &   0.00160   &    0.00117   &    0.00114   &   II   &   H\"ubscher et al. (2005)      \\
3,221.4928   &   17998.0   &   0.00174   &    0.00131   &    0.00128   &   I    &   H\"ubscher et al. (2005)      \\
3,224.4358   &   18008.5   &   0.00035   &   -0.00008   &   -0.00010   &   II   &   Pribulla et al. (2005)        \\
3,226.5392   &   18016.0   &   0.00062   &    0.00019   &    0.00018   &   I    &   Pribulla et al. (2005)        \\
3,226.5392   &   18016.0   &   0.00062   &    0.00019   &    0.00018   &   I    &   H\"ubscher et al. (2005)      \\
3,228.3612   &   18022.5   &  -0.00010   &   -0.00052   &   -0.00053   &   II   &   Br\'at et al. (2007)          \\
3,233.4095   &   18040.5   &   0.00068   &    0.00026   &    0.00026   &   II   &   H\"ubscher et al. (2005)      \\
3,233.5488   &   18041.0   &  -0.00023   &   -0.00065   &   -0.00065   &   I    &   H\"ubscher et al. (2005)      \\
3,236.3538   &   18051.0   &   0.00059   &    0.00018   &    0.00019   &   I    &   Pribulla et al. (2005)        \\
3,236.4931   &   18051.5   &  -0.00032   &   -0.00073   &   -0.00072   &   II   &   Pribulla et al. (2005)        \\
3,240.4180   &   18065.5   &  -0.00127   &   -0.00168   &   -0.00166   &   II   &   Pribulla et al. (2005)        \\
3,240.5597   &   18066.0   &   0.00023   &   -0.00019   &   -0.00017   &   I    &   Pribulla et al. (2005)        \\
3,250.3747   &   18101.0   &   0.00060   &    0.00020   &    0.00024   &   I    &   H\"ubscher et al. (2005)      \\
3,250.5140   &   18101.5   &  -0.00031   &   -0.00071   &   -0.00067   &   II   &   H\"ubscher et al. (2005)      \\
3,255.4219   &   18119.0   &   0.00028   &   -0.00011   &   -0.00006   &   I    &   H\"ubscher et al. (2005)      \\
3,255.5628   &   18119.5   &   0.00097   &    0.00058   &    0.00063   &   II   &   H\"ubscher et al. (2005)      \\
3,257.3845   &   18126.0   &  -0.00004   &   -0.00044   &   -0.00038   &   I    &   H\"ubscher et al. (2005)      \\
3,257.5226   &   18126.5   &  -0.00215   &   -0.00254   &   -0.00249   &   II   &   H\"ubscher et al. (2005)      \\
3,282.3417   &   18215.0   &  -0.00003   &   -0.00039   &   -0.00029   &   I    &   H\"ubscher et al. (2005)      \\
3,282.4834   &   18215.5   &   0.00146   &    0.00110   &    0.00120   &   II   &   H\"ubscher et al. (2005)      \\
3,341.2285   &   18425.0   &  -0.00097   &   -0.00126   &   -0.00106   &   I    &   Diethelm (2005)               \\
3,360.2974   &   18493.0   &  -0.00048   &   -0.00075   &   -0.00053   &   I    &   Zejda et al. (2006)           \\
3,601.4569   &   19353.0   &  -0.00027   &   -0.00027   &    0.00010   &   I    &   H\"ubscher et al. (2006)      \\
3,613.3741   &   19395.5   &  -0.00083   &   -0.00081   &   -0.00044   &   II   &   H\"ubscher et al. (2006)      \\
3,613.5150   &   19396.0   &  -0.00013   &   -0.00012   &    0.00025   &   I    &   Zejda et al. (2006)           \\
3,613.5159   &   19396.0   &   0.00077   &    0.00078   &    0.00115   &   I    &   H\"ubscher et al. (2006)      \\
3,614.3566   &   19399.0   &   0.00021   &    0.00023   &    0.00059   &   I    &   Br\'at et al. (2007)          \\
3,614.4970   &   19399.5   &   0.00040   &    0.00042   &    0.00079   &   II   &   H\"ubscher et al. (2006)      \\
3,616.4573   &   19406.5   &  -0.00222   &   -0.00220   &   -0.00184   &   II   &   Br\'at et al. (2007)          \\
3,617.4407   &   19410.0   &  -0.00028   &   -0.00026   &    0.00010   &   I    &   Br\'at et al. (2007)          \\
3,632.0216   &   19462.0   &  -0.00111   &   -0.00107   &   -0.00071   &   I    &   Nagai (2006)                  \\
3,632.1623   &   19462.5   &  -0.00062   &   -0.00058   &   -0.00021   &   II   &   Nagai (2006)                  \\
3,648.4272   &   19520.5   &   0.00005   &    0.00011   &    0.00047   &   II   &   H\"ubscher et al. (2006)      \\
3,651.3711   &   19531.0   &  -0.00043   &   -0.00038   &   -0.00001   &   I    &   Br\'at et al. (2007)          \\
3,651.3714   &   19531.0   &  -0.00013   &   -0.00008   &    0.00029   &   I    &   H\"ubscher et al. (2006)      \\
3,651.5111   &   19531.5   &  -0.00064   &   -0.00059   &   -0.00022   &   II   &   H\"ubscher et al. (2006)      \\
3,659.3629   &   19559.5   &  -0.00054   &   -0.00047   &   -0.00011   &   II   &   H\"ubscher et al. (2006)      \\
3,659.5033   &   19560.0   &  -0.00035   &   -0.00028   &    0.00008   &   I    &   H\"ubscher et al. (2006)      \\
3,663.9891   &   19576.0   &  -0.00123   &   -0.00116   &   -0.00080   &   I    &   Nagai (2006)                  \\
3,951.1360   &   20600.0   &  -0.00209   &   -0.00177   &   -0.00152   &   I    &   Nagai (2007)                  \\
3,951.2770   &   20600.5   &  -0.00129   &   -0.00098   &   -0.00073   &   II   &   Nagai (2007)                  \\
3,966.4203   &   20654.5   &  -0.00055   &   -0.00022   &    0.00002   &   II   &   H\"ubscher \& Walter (2007)   \\
3,966.5597   &   20655.0   &  -0.00136   &   -0.00103   &   -0.00079   &   I    &   H\"ubscher \& Walter (2007)   \\
3,985.4890   &   20722.5   &  -0.00025   &    0.00009   &    0.00031   &   II   &   Do\u{g}ru et al. (2007)       \\
3,989.6960   &   20737.5   &   0.00048   &    0.00082   &    0.00105   &   II   &   Ogloza et al. (2008)          \\
3,992.3574   &   20747.0   &  -0.00209   &   -0.00174   &   -0.00152   &   I    &   H\"ubscher \& Walter (2007)   \\
4,000.3487   &   20775.5   &  -0.00269   &   -0.00234   &   -0.00213   &   II   &   Br\'at et al. (2007)          \\
4,002.4524   &   20783.0   &  -0.00212   &   -0.00178   &   -0.00156   &   I    &   H\"ubscher \& Walter (2007)   \\
4,279.647    &   21771.5   &  -0.00040   &    0.00003   &    0.00008   &   II   &   Paschke (2007)                \\
4,327.4581   &   21942.0   &  -0.00050   &   -0.00008   &   -0.00006   &   I    &   Br\'at et al. (2007)          \\
4,328.4386   &   21945.5   &  -0.00147   &   -0.00105   &   -0.00102   &   II   &   Br\'at et al. (2007)          \\
4,328.4386   &   21945.5   &  -0.00147   &   -0.00105   &   -0.00102   &   II   &   Br\'at et al. (2007)          \\
4,328.4393   &   21945.5   &  -0.00077   &   -0.00035   &   -0.00032   &   II   &   Br\'at et al. (2007)          \\
4,330.5428   &   21953.0   &  -0.00040   &    0.00002   &    0.00004   &   I    &   Br\'at et al. (2007)          \\
4,330.5431   &   21953.0   &  -0.00010   &    0.00032   &    0.00034   &   I    &   Br\'at et al. (2007)          \\
4,359.9874   &   22058.0   &   0.00035   &    0.00076   &    0.00077   &   I    &   Nagai (2008)                  \\
4,386.6281   &   22153.0   &   0.00138   &    0.00178   &    0.00178   &   I    &   Samolyk (2008a)               \\
4,420.5567   &   22274.0   &  -0.00056   &   -0.00017   &   -0.00018   &   I    &   Samolyk (2008a)               \\
4,650.7799   &   23095.0   &  -0.00021   &    0.00002   &   -0.00006   &   I    &   Samolyk (2008b)               \\
4,702.6579   &   23280.0   &   0.00054   &    0.00071   &    0.00062   &   I    &   Samolyk (2008b)               \\
4,736.72819  &   23401.5   &   0.00009   &    0.00022   &    0.00014   &   II   &   this article                  \\
4,736.86805  &   23402.0   &  -0.00026   &   -0.00012   &   -0.00021   &   I    &   this article                  \\
4,737.70921  &   23405.0   &  -0.00035   &   -0.00022   &   -0.00031   &   I    &   this article                  \\
4,737.84996  &   23405.5   &   0.00019   &    0.00032   &    0.00024   &   II   &   this article                  \\
4,738.69117  &   23408.5   &   0.00015   &    0.00028   &    0.00019   &   II   &   this article                  \\
4,759.72254  &   23483.5   &   0.00020   &    0.00030   &    0.00022   &   II   &   this article                  \\
4,760.70325  &   23487.0   &  -0.00055   &   -0.00045   &   -0.00053   &   I    &   this article                  \\
4,761.68556  &   23490.5   &   0.00030   &    0.00040   &    0.00031   &   II   &   this article                  \\
4,785.66083  &   23576.0   &  -0.00013   &   -0.00006   &   -0.00015   &   I    &   this article                  \\
4,786.64256  &   23579.5   &   0.00014   &    0.00020   &    0.00012   &   II   &   this article                  \\
\enddata
\end{deluxetable}

\begin{deluxetable}{cccc}
\tablewidth{0pt}
\tablecaption{The fitting parameters for ephemeris forms (1) and (2) for BX Peg.$\rm ^a$}
\tablehead{
\colhead{Parameter}     &  \colhead{Equation (1)}         &  \colhead{Equation (2)}         &  \colhead{Unit}
}                                                                                         
\startdata                                                                                
$T_0$                   &  2,448,174.52857(31)            &  2,448,174.52814(23)            &  HJD             \\
$P$                     &  0.280419296(18)                &  0.280419347(13)                &  d               \\
$a_{12}\sin i_{3}$      &  2.464(75)                      &  2.458(50)                      &  AU              \\
$\omega$                &  148.9(3.0)                     &  141.4(2.1)                     &  deg             \\
$e$                     &  0.715(36)                      &  0.627(38)                      &                  \\
$n$                     &  0.016535(92)                   &  0.018422(73)                   &  deg d$^{-1}$    \\
$T$                     &  2,425,845(123)                 &  2,427,824(80)                  &  HJD             \\
$P_{3}$                 &  59.61(33)                      &  53.50(21)                      &  yr              \\
$K$                     &  0.01126(34)                    &  0.01238(25)                    &  d               \\
$f(M_{3})$              &  0.00421(13)                    &  0.00519(11)                    &  $M_\odot$       \\
$M_3 \sin i_{3}$        &  0.223(4)                       &  0.241(2)                       &  $M_\odot$       \\[2.0mm]
$A$                     &  $-$3.944(77)$\times 10^{-11}$  &  $-$3.736(56)$\times 10^{-11}$  &  d               \\
$dP$/$dt$               &  $-$10.28(20)$\times 10^{-8}$   &  $-$9.73(15)$\times 10^{-8}$    &  d yr$^{-1}$     \\[2.0mm]
$K^{\prime}$            &  \dots                          &  0.00124(31)                    &  d               \\
$\omega^{\prime}$       &  \dots                          &  0.0227(10)                     &  deg $P^{-1}$    \\
$\omega$$_0 ^{\prime}$  &  \dots                          &  112(16)                        &  deg             \\
$P^{\prime}$            &  \dots                          &  12.2(1.9)                      &  yr              \\
\enddata
\tablenotetext{a}{A parenthesized number is the 1$\sigma$-value of the last two digits of each parameter.}
\end{deluxetable}

\begin{deluxetable}{cccc}
\tablewidth{0pt}
\tablecaption{Parameters for the quadratic {\it plus} two-LTT ephemeris of BX Peg.}
\tablehead{
\colhead{Parameter}      & \colhead{Third body}       & \colhead{Fourth body}     & \colhead{Unit}   \\
\colhead{}               & \colhead{$\tau_{3}$}       & \colhead{$\tau_{4}$}      &                    
}
\startdata
$T_0$                    &  \multicolumn{2}{c}{2,448,174.52786(19)}               &   HJD            \\
$P$                      &  \multicolumn{2}{c}{0.280419354(10)}                   &   d              \\
$a_{12}\sin i_{3,4}$     &  2.342(41)                 &  0.418(55)                &   AU             \\
$\omega$                 &  166.8(1.7)                &  139.7(7.5)               &   deg            \\
$e$                      &  0.488(27)                 &  0.723(104)               &                  \\
$n$                      &  0.017874(76)              &  0.06206(19)              &   deg d$^{-1}$   \\
$T$                      &  2,428,381(89)             &  2,423,967(83)            &   HJD            \\
$P_{3,4}$                &  55.15(23)                 &  15.88(5)                 &   yr             \\
$K$                      &  0.01190(21)               &  0.00201(26)              &   d              \\
$f(M_{3,4})$             &  0.004227(76)              &  0.000289(38)             &   M$_\odot$      \\
$M_{3,4} \sin i_{3,4}$   &  0.223(2)                  &  0.107(5)                 &   M$_\odot$      \\[2.0mm]
$A$                      &  \multicolumn{2}{c}{$-$3.778(46)$\times 10^{-11}$}     &   d              \\
$dP$/$dt$                &  \multicolumn{2}{c}{$-$9.84(12)$\times 10^{-8}$}       &   d yr$^{-1}$    \\
\enddata
\end{deluxetable}

\begin{deluxetable}{cccc}
\tablewidth{0pt}
\tablecaption{Applegate parameters for the 12-yr period modulation of BX Peg.}
\tablehead{
\colhead{Parameter}       & \colhead{Primary}      & \colhead{Secondary}     & \colhead{Unit}
}
\startdata
$\Delta P$                & 0.0410                 &  0.0410                 &  s                   \\
$\Delta P/P$              & $1.69\times10^{-6}$    &  $1.69\times10^{-6}$    &                      \\
$\Delta Q$                & ${2.81\times10^{48}}$  &  ${7.54\times10^{48}}$  &  g cm$^2$            \\
$\Delta J$                & ${2.01\times10^{46}}$  &  ${3.89\times10^{46}}$  &  g cm$^{2}$ s$^{-1}$ \\
$I_{\rm s}$               & ${9.48\times10^{52}}$  &  ${6.11\times10^{53}}$  &  g cm$^{2}$          \\
$\Delta \Omega$           & ${2.12\times10^{-7}}$  &  ${6.35\times10^{-8}}$  &  s$^{-1}$            \\
$\Delta \Omega / \Omega$  & ${8.18\times10^{-4}}$  &  ${2.45\times10^{-4}}$  &                      \\
$\Delta E$                & ${8.53\times10^{39}}$  &  ${4.94\times10^{39}}$  &  erg                 \\
$\Delta L_{\rm rms}$      & ${6.96\times10^{31}}$  &  ${4.03\times10^{31}}$  &  erg s$^{-1}$        \\
                          & 0.018                  &  0.010                  &  $L_\odot$           \\
                          & 0.055                  &  0.016                  &  $L_{\rm p,s}$       \\
$\Delta m_{\rm rms}$      & $\pm$0.020             &  $\pm$0.011             &  mag                 \\
$B$                       & 11.2                   &  8.1                    &  kG
\enddata
\end{deluxetable}

\begin{deluxetable}{ccccccccccc}
\tablewidth{0pt}
\tablecaption{Light levels of BX Peg at four different phases.}
\tablehead{
Mean Year & Mean Epoch   & \multicolumn{2}{c}{Min I} & \multicolumn{2}{c}{Max I} & \multicolumn{2}{c}{Min II} & \multicolumn{2}{c}{Max II} & Ref. \\
          &              & $\Delta V$ & $\Delta B$   & $\Delta V$ & $\Delta B$   & $\Delta V$ & $\Delta B$    &  $\Delta V$ & $\Delta B$   &  
}
\startdata
 1978.84  &  -15537.50   & 0.637 & 0.348  & 0.009 & -0.309  & 0.612 & 0.330  & 0.011 & -0.320  & 1 \\
 1988.63  &   -2787.62   & 0.736 & 0.406  & 0.127 & -0.251  & 0.709 & 0.367  & 0.104 & -0.278  & 2 \\
 1999.83  &   11807.24   & 0.785 & 0.498  & 0.157 & -0.160  & 0.749 & 0.455  & 0.125 & -0.198  & 3 \\
 2000.67  &   12897.27   & 0.776 & 0.486  & 0.133 & -0.197  & 0.700 & 0.410  & 0.112 & -0.191  & 3 \\
 2008.81  &   23490.45   & 0.890 & 0.622  & 0.226 & -0.076  & 0.765 & 0.495  & 0.207 & -0.094  & 4 \\
          &              & 0.765 & 0.472  & 0.130 & -0.199  & 0.707 & 0.411  & 0.112 & -0.216  & 5 \\
\enddata
\tablerefs{(1) Zhai \& Zhang (1979); (2) Samec (1990); (3) Lee04a; (4) this article; (5) mean value.}
\end{deluxetable}

\end{document}